\documentclass[showpacs,final,a4paper,pra,10pt,twocolumn]{revtex4}
\usepackage{amsmath}

\input{tcilatex}

\begin{document}

\title{Conditional transfer of quantum correlation in the intensity of twin
beams}
\author{Yun Zhang\thanks{%
Email: zhangyun@nict.go.jp}, Kazuhiro Hayasaka and Katsuyuki Kasai}
\affiliation{Kansai Advanced Research Center, National Institute of Information and
Communications Technology, 588-2 Iwaoka, Nishi-ku, Kobe, 651-2492, Japan}
\date{\today }

\begin{abstract}
A conditional protocol of transferring quantum-correlation in continuous
variable regime was experimentally demonstrated. The quantum-correlation in
two pairs of twin beams, each characterized by intensity-difference
squeezing of $7.0\pm 0.3$ dB, was transferred to two initially independent
idler beams. The quantum-correlation transfer resulted in
intensity-difference squeezing of $4.0\pm 0.2$ dB between two idler beams.
The dependence of preparation probability and transfer fidellity on the
selection bandwidth was also studied.
\end{abstract}

\maketitle

The ability to transfer quantum properties from one system to another is a
prerequisite for quantum communications. A well-known example is the quantum
state transfer between atoms and photons. Unitary interactions between atoms
and photons, such as strong-coupling cavity QED \cite{kimble} and
electromagnetically induced transparency \cite{liu}, have been employed to
transfer an initial quantum state from one subsystem to another subsystem.
Quantum frequency conversion is another example, in which the quantum
properties of an optical wavelegth is transferred to another \cite{kumar1}.
More popular example is the quantum teleportation \cite{furusawa}. The
initial unknown state of a quantum system can be transferred to another with
the assistance of the entanglement. All of these protocols are unconditional
and demand a lot of effort for experimental realization. Here we demonstrate
a simpler protocol, in which quantum correlation of twin beams is
conditionally transferred from the one pair to another pair by postselection.

Differing from squeezed states on quadrature amplitudes, the intensity
difference noise between signal and idler beams of twin beams is below the
shot noise limit, therefore only the field intensities need to be measured
rather than the quadrature amplitudes. Since the first experimental
demonstration of twin beams, their generation and application have been
studied extensively \cite{heidmann}. Recent investigations have focused on
the characterizing twin beams in the time domain \cite{zhang1} and
conditional preparation of a sub-Poissonian state based on post selection of
twin beams \cite{fabre1}. A more exciting achievement is that ultra-stable
twin beams have been produced at a precise frequency degeneracy from a
nondegenerate optical parametric oscillator (NOPO) operating above its
threshold through the use of a new nonlinear crystal and an active
phase-lock system \cite{feng}. The entanglement state can also be generated
by using a self-phase-locked OPO \cite{fabre2}. These reports indicate that
the quantum correlation between the two beams of twin beams can be as a
testbed for verifying the foundations of the quantum theory in quantum
optics and quantum information fields.

Conditional state preparation, or post selection, was originally proposed
for use in a discrete-variable system. It was widely and successfully used
to prepare a single photon state from quantum-correlated photons
(\textquotedblleft twin photons\textquotedblright ) generated via parametric
down conversion \cite{lvovsky}. Quantum correlation between two photons - a
signal and a trigger - is the prerequisite for such a general procedure.
When a single photon detector, located in one of the emission channels,
registers a single photon, the correlated twin-photon state collapses into a
single photon in a well-defined spatiotemporal mode traveling along the
other emission channel. This technique has been used in many experiments,
since it was originally proposed. State collapse is obviously not restricted
to the case of photon counting, so it may be interesting to extend this
technique to a continuous-variable regime. Continuous detection conditioned
by a photon counting event has been implemented in various schemes, such as
generation of a \textquotedblleft degaussification\textquotedblright\ state 
\cite{grangier}, tests of quantum nonlocality \cite{nha}, entanglement
purification\cite{opa} and generation of a Schr\H{o}dinger-cat state. In a
cavity QED experiment, conditional measurements of atomic states have also
led to the experimental generation of optical states with a well-defined
phase \cite{foster}. On the other hand, Lvovsky et al. prepared a bit of
quantum information encoded in a discrete basis conditioned on observation
of a continuous observable \cite{lvovsky2}. The conditional preparation of a
nonclassical state of light was also experimentally demonstrated in
continuous-variable regime \cite{fabre2, fabre3}. So far, discrete- and
continuous- variable quantum information science has developed with some
overlap between these two domains. To the best of our knowledge, no scheme
has been suggested to transfer quantum properties by technique of
conditional preparation in the continuous variables. This is the purpose of
the present letter, in which the degree of quantum correlation $4.0\pm 0.2$
dB below the shot-noise limit for the intensity difference resulting from
two pairs of twin beams exhibiting 7.0 dB is transferred using the simple
conditional preparation technique.

\FRAME{ftbpFU}{3.5578in}{3.5094in}{0pt}{\Qcb{Schematic of experimental
setup. OPO, optical parametric oscillator; M mirror; P, half-wave plate;
PBS, polarizing beam splitter; BS, beam splitter; D, photodiodes; LPF,
low-pass filter; and G, low-noise electronic amplifier.}}{}{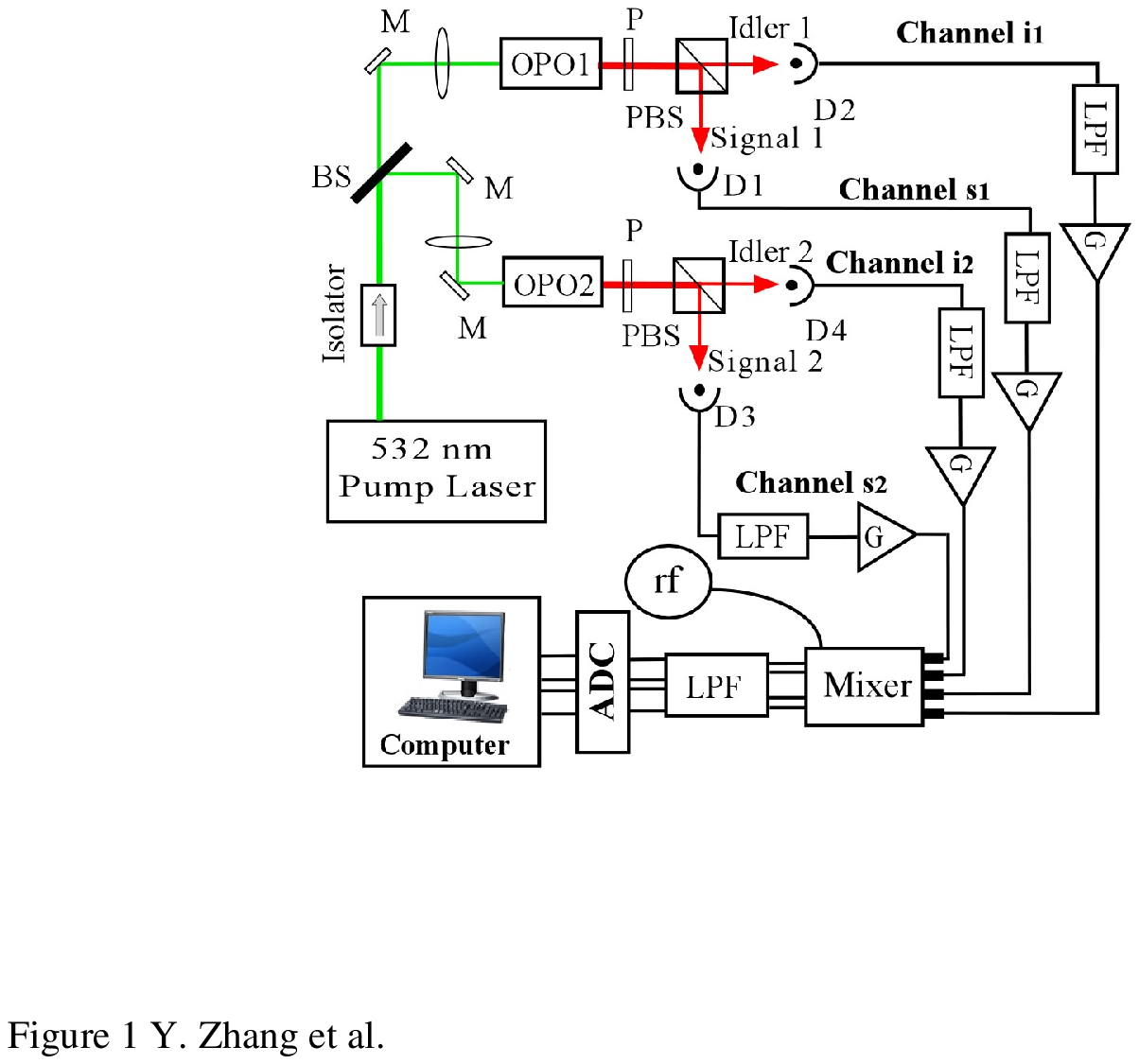}{\special%
{language "Scientific Word";type "GRAPHIC";maintain-aspect-ratio
TRUE;display "USEDEF";valid_file "F";width 3.5578in;height 3.5094in;depth
0pt;original-width 8.188in;original-height 10.562in;cropleft
"0.2819";croptop "0.7933";cropright "0.7130";cropbottom "0.4636";filename
'fig1.ps';file-properties "XNPEU";}}The experimental setup is outlined in
Fig. 1. Two independent quantum-correlated twin beams at 1064 nm were
produced by two triply resonant NOPOs. The NOPOs were pumped by a
diode-pumped cw frequency-doubled Nd:YAG laser. To increase mechanical
stability and reduce extra loss, both the OPO1 and the OPO2 have a
semi-monolithic configuration. Each of them consists of a 10 mm-long KTP
crystal and concave mirror with a curvature of 20 mm. The input coupler is
directly coated on one face of the KTP crystal. The reflectivity of the
input coupler is 90\% for the pump (532 nm) and high reflectivity for the
signal and idler beams (1064 nm). The output coupler is high reflectivity
for the pump and its transmission is 5\% for the infrared. OPOs lengths are
actively locked on the pump resonance. For this purpose, the piezoelectric
transducer (PZT), on which the output coupler is mounted, is dithered at 100
kHz and the pump light leaking from the cavity is monitored by a photodiode.
The error signal was derived by mixing the photocurrent and another 100 kHz
rf signal. The error signal is fed back to the PZT after passing through a
home-made proportional-integral controller. This kept the cavity length to
the triply-resonant frequency. At exact triple resonance, the oscillation
threshold is less than 10 mW. In particular, at a pump power of 30 mW an
output power of 16 mW was obtained. The pump power of the OPOs was adjusted
so that the two output twin beams were of equal intensity. The configuration
and operating principles of the NOPOs have been described in detail
previously \cite{zhang1,zhang3}. Both produce twin beams of 7.0 dB of
intensity difference squeezing at 3.5 MHz as photocurrent difference
measured using a spectrum analyzer. The two pairs of orthogonally polarized
twin beams from OPO1 and OPO2 were divided using two polarizing beam
splitters. The output of the PBSs were directly detected using four balanced
high quantum efficiency photodiodes. A half-wave plate was inserted before
the polarizing beam splitters. It was used to rotate the polarization of the
twin beams, which enabled us to measure the shot-noise level \cite{heidmann}%
. When the polarization of twin beams are rotated by an angle of 0$^{\circ }$
to the PBS axis, we record the quantum correlation between twin signal and
idler beams of twin beams. However, when it is rotated by an angle of 45$%
^{\circ }$, we can record the shot-noise level. To reconfirm the shot-noise
level, a coherent light with the same power of the twin beams was input to
the other ports of PBS1 and PBS2 (not shown in Fig. 1).

The detection and implementation of the quantum-correlation transfer system
was done in the time domain, not in the frequency domain. We accessed the
full quantum characterization of the twin beams at a given Fourier frequency 
$\Omega .$ Each of photocurrent detected by the photodiodes (D1 to D4) was
amplified by a low-noise 46-dB gain amplifier after it passed through a
21.4-MHz low-pass filter. The photocurrent was then mixed with an electrical
local oscillator of frequency $\Omega $. The near-dc region downconverted
frequency output of the mixer was further amplified and filtered using a
100-kHz low-pass filter to prevent the signal from being averaged over a
large frequency range. It was then digitized at a sample rate of 200 kHz
using a 12-bit, 4-channel acquisition card. To establish the
quantum-correlation between the two idler lights of two twin beams (channel
i1 and channel i2), we simultaneously\ measured the intensity of the two
signal lights of the twin beams. Upon measurement, the two idler beams of
the twin beams will collapse into an eigenstate of this observable \cite%
{fabre1}. When the measurement results of two signal beams have the same
values, the two idler state will collapse into the same state. In other
words, as we demonstrate below, quantum correlation can be established
between the two initially independent idler beams of the two pairs of twin
beams if we post-select the events based on the measured results of two
signal beams of the two pairs of twin beams (channels s1 and s2).

In the experiments, we first locked both OPO1 and OPO2 on the pump
resonance. The OPOs operated stably for more than half an hour without mode
hopping. The generated twin beams were measured using a spectrum analyzer in
the frequency domain \cite{heidmann} or by mixing the photocurrent with a
sinusoidal local oscillator in the time domain \cite{zhang1,fabre1}. Both
measurement results indicated quantum correlation of $7.0\pm 0.3$ dB between
the signal and idler beams, which were output from both OPO1 or OPO2.\
Transfer of the quantum correlation between the two idler beams of two pairs
of the twin beams was then performed. For each input state, which
corresponded to coherent light and twin beams rotated by 45$^{\circ }$ and 0$%
^{\circ },$ two successive acquisitions (300,000 points for each channel)
were acquired, one under conditional selection and the other under
unconditional selection.

It is well-known that twin beams can be characterized by their joint photon
number distribution \cite{walls,kumar}. In the ideal case of a system with
no losses, the joint photon number distribution is $p(n_{s},n_{i})=0$, when $%
n_{s}\neq n_{i}$; however $p(n_{s},n_{i})\neq 0$, when $n_{s}=n_{i}$. This
means that the idler beam of twin beams will collapse into a perfect
N-photon Fock state when an N-photon state\ is detected at the signal beam.
Now, supposing a system consists of two pairs of twin beams, in other words,
it has joint photon number distributions $p_{1}(n_{s1},n_{i1})$ and $%
p_{2}(n_{s2},n_{i2})$, which have the above-mentioned characteristics; the
two idler beams will simultaneously collapse into an N-photon state when
N-photon states are detected at two signal beams of two pairs of twin beams
at the same time, i. e. $n_{s1}=n_{s2}=N$. Therefore, quantum correlation
between the two idler beams will be automatically established and the photon
number distribution $p_{3}(n_{i1},n_{i2})$\ between them will have the
above-mentioned characteristics of twin beams. This shows that quantum
correlation can be transferred from the system $(n_{s1(2)},n_{i1(2)})$\ to
the system $(n_{i1},n_{i2})$.

In a real experiment, the correlation between the signal and idler beams,
which are generated by an OPO, is not perfect, and the Fourier components of
the signal and idler intensity quantum fluctuations are correlated only when
they lie inside the cavity bandwidth. The imperfect correlation and
less-than unity detection efficiency results in the correlation being
characterized by the limited intensity difference squeezing. In other words,
the measured photon number difference fluctuation distribution $%
p(n_{s}-n_{i})$ between the signal and idler of twin beams will be narrower
than that of two uncorrelated beams \cite{zhang1}. The instantaneous values
of the signal and idler photocurrents therefore play a role in the
occurrence of counts in a photon counting regime. The conditional
preparation is that we select the idler photocurrent only during time
intervals when the signal photocurrent satisfies the post-selection
conditions (see also \cite{fabre1,lam}).

The post-selection process is performed as follows: we first compare the
results of measuring the intensity of the two signal beams ($I_{s1}$, $I_{s2}
$). The intensity values of the two idler beams ($I_{i1}$, $I_{i2}$) are
kept only if the difference between intensity values ($I_{s1}-I_{s2}$) have
a value of zero (within a band $\Delta I$ smaller than the photocurrent
standard deviation), which means we measured the same state at two signal
beams. As we will show, this selection provides quantum-correlation transfer
using two pairs of twin beams.

\FRAME{ftbpFU}{3.5846in}{1.8637in}{0pt}{\Qcb{Experimental results: (a)
correlation between two idler intensity fluctuations (only 20 000 shown).
Cyan points correspond unconditioned results with input state of twin beams
rotated by 45$^{\circ }$ and 0$^{\circ }$; green points correspond to
conditioned results with input state of twin beams rotated by 45$^{\circ }$,
indicate experimentally measured shot noise level; red points correspond to
conditioned and unconditioned results with input state of coherent state,
indicate the reconfirmed shot noise level; and blue points correspond to
conditioned results of input state with twin beams rotated by 0$^{\circ }$,
the narrowing distribution of the points indicates the sucessfully
transferred quantum correlation. (b) The conditioned and unconditioned
photocurrent difference probability with various input states. Selection
bandwidth equal $0.03\protect\delta $.}}{}{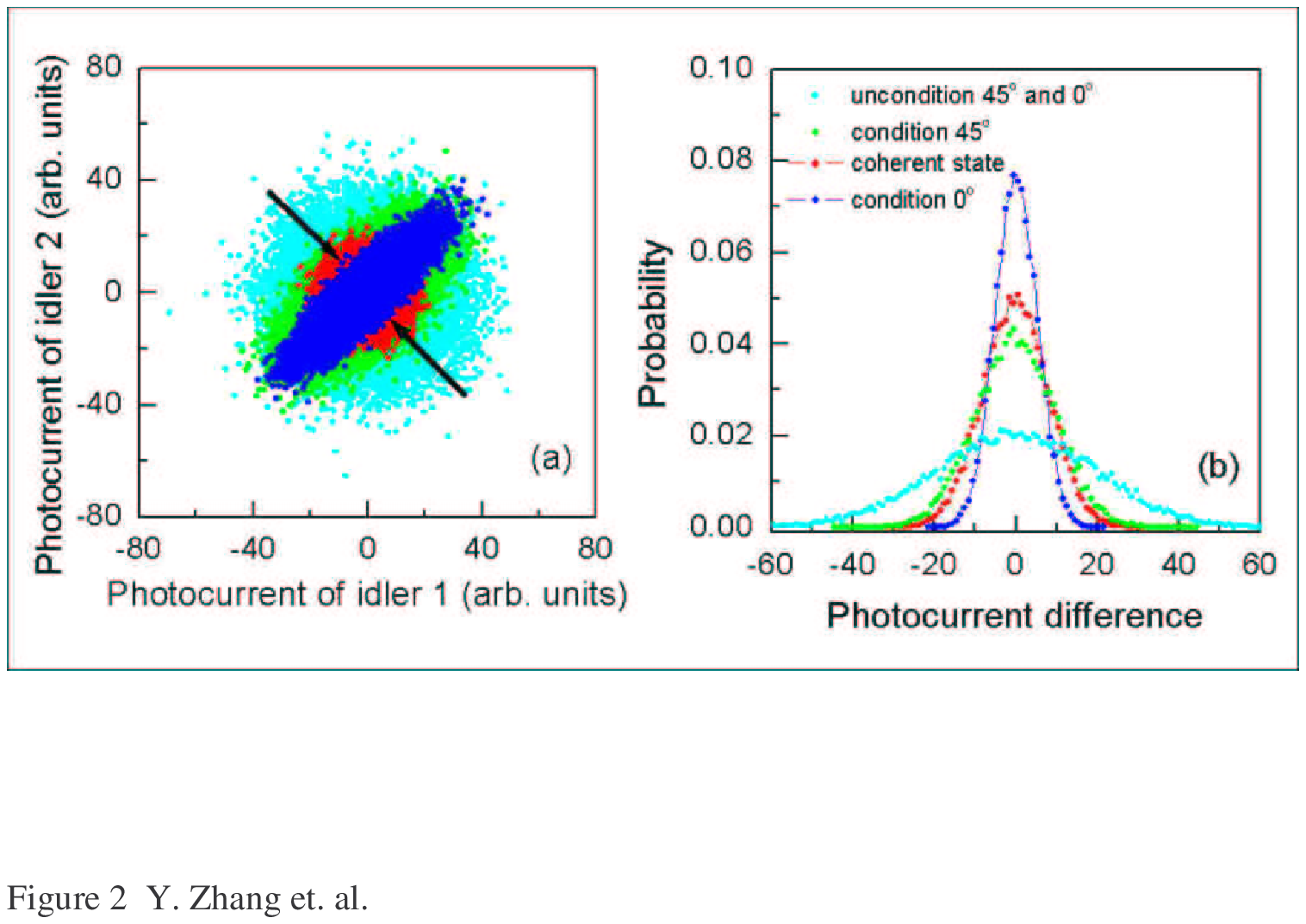}{\special{language
"Scientific Word";type "GRAPHIC";maintain-aspect-ratio TRUE;display
"USEDEF";valid_file "F";width 3.5846in;height 1.8637in;depth
0pt;original-width 7.7574in;original-height 11.0627in;cropleft
"0.1527";croptop "0.7699";cropright "0.8693";cropbottom "0.5105";filename
'fig2.ps';file-properties "XNPEU";}}

Figure 2 sums up the measurements obtained at a local frequency of $\Omega
=3.5$ MHz. Figure 2 (a) shows the actual recording. Each point corresponds
to one simultaneous measurement of the photocurrent fluctuations of two
idler beams of two pairs of twin beams. The direction of arrows indicates
the correlation between two beams. It is also showed that the recorded
photocurrent, when the polarization of twin beams are rotated an angle of 45$%
^{\circ }$, can be as the shot noise level. As the OPO was pumped above its
threshold, the idler and signal beams produced intensity fluctuations, which
were much larger than the shot noise level \cite{fabre1,zhang1}. As expected
for twin beams rotated by 0$^{\circ }$, a quantum correlation between the
two idler beams was established when we post selected the events based on
the measured result of the two signal beams of the two pairs of twin beams.
To quantify the amount of measured quantum correlation between the two idler
beams, we determined the variances of the stored dats in the photocurrent
difference of twin beams normalized to that of coherent state. We calculated
the noise variance in the difference between the photocurrent fluctuation of
the two idler beams. It reached a value of $4.0\pm 0.2$ dB below the
shot-noise level \cite{zhang1}. To show the quantum character of the
measured distribution, we give the probability distribution in Fig. 2 (b). A
histograms or probability distribution can be constructed from the data in
Fig. 2 (a). In the case of conditional and unconditional selection, the
probability distributions of photocurrent difference between the channels i1
and i2 corresponding to the coherent state, twin beams rotated by 45$^{\circ
}$ and 0$^{\circ }$ as input states, are plotted on Fig. 2 (b). They were
obtained by binning up the data in Fig. 2 (a) according to the value of
photocurrent differences between two two channels with a bin size much
smaller than that of the standard deviation $\delta $ of a coherent state
with the same power \cite{fabre1,zhang1}. The measured distribution of
photocurrents difference between two idler beams is narrower than that of
two coherent states, indicating the realization of quantum-correlation
transfer between the two pairs of twin beams.

\FRAME{ftbpFU}{3.5025in}{2.2364in}{0pt}{\Qcb{Measured transferred
quantum-correlation as a function of different values of correlation of twin
beams (a) and selection bandwidth on trigger light normalized to $\protect%
\delta $ (b). In figure (a), selection bandwidth is taken to equal $0.03%
\protect\delta $. Squares: experimental data.}}{}{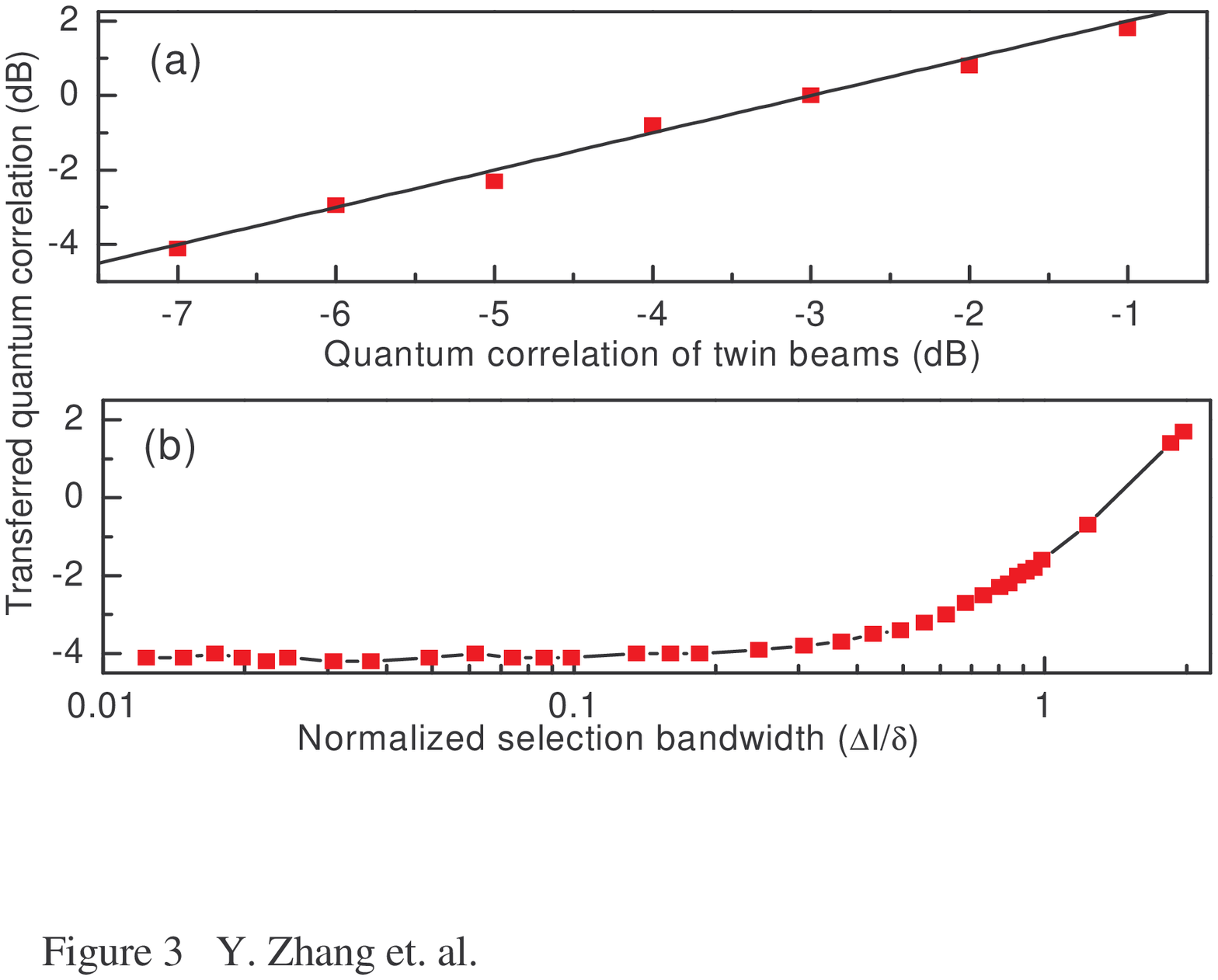}{\special{language
"Scientific Word";type "GRAPHIC";maintain-aspect-ratio TRUE;display
"USEDEF";valid_file "F";width 3.5025in;height 2.2364in;depth
0pt;original-width 7.7574in;original-height 11.0627in;cropleft
"0.0506";croptop "0.7332";cropright "0.9466";cropbottom "0.3338";filename
'fig3.ps';file-properties "XNPEU";}}

Just as the same as the case of conditional preparation of a sub-Poissonian
state from twin beams \cite{fabre1,fabre2}, the transfer fidelity and
preparation probability strongly depend on the selection bandwidth ($\Delta I
$). The success rate can be improved by increasing the selection bandwidth
at the expense of a decrease in the transferred quantum correlation degree
(transfer fidelity). In Fig. 3, we give the measured transferred quantum
correlation in the conditionally produced transfer as a function of
different amounts of quantum correlations of the twin beams (Fig. 3a), which
can be varied by inserting losses on the OPO beams or rotating the half-wave
plate (P), and the selection bandwidth normalized to $\delta $ (Fig. 3b).
All the measured results show that quantum correlation between two idler
beams of two independent twin beams was established when the quantum
correlation of the twin beams was larger than 3 dB. The fidelity of
transferred quantum correlation was degraded by 3 dB from the initial
quantum correlation of twin beams \cite{fabre1}. In the range where $\Delta
I/\delta $ is very small, the quantum correlation transfer is almost
constant until the normalized selection bandwidth reaches the order of $0.1$%
. In Fig. 4, we give the preparation probability for different normalized
selection bandwidth (Fig. 4a) and different amounts of quantum correlation
of the twin beams (Fig. 4b). We note that the preparation probability in our
experiment is different from that in the experiment of sub-Poissonian state
generation from twin beams \cite{fabre1}, because we selected the events
based on the classical coincidence between two original twin beams differing
from events which are kept only if signal and idler photocurrents values
simultaneously\ fall inside a narrow band around the preselected mean value
in Ref. 7. In other words, events of idler intensity values around various
mean values are selected in our experiment, however,\ events of idler
intensity values around an unitary mean value were selected in
sub-Poissonian state generation experiment. Due to the postselection based
on the classical coincidence between two independent twin beams, the
preparation probability is insensitive to noise reduction of the twin beams.

\FRAME{ftbpFU}{3.5042in}{2.418in}{0pt}{\Qcb{Preparation probability for
different normalized selection bandwidth (a) and different values of
correlation of twin beams (b). In figure (b), selection bandwidth is taken
to equal $0.03\protect\delta $. Circles: experimental data.}}{}{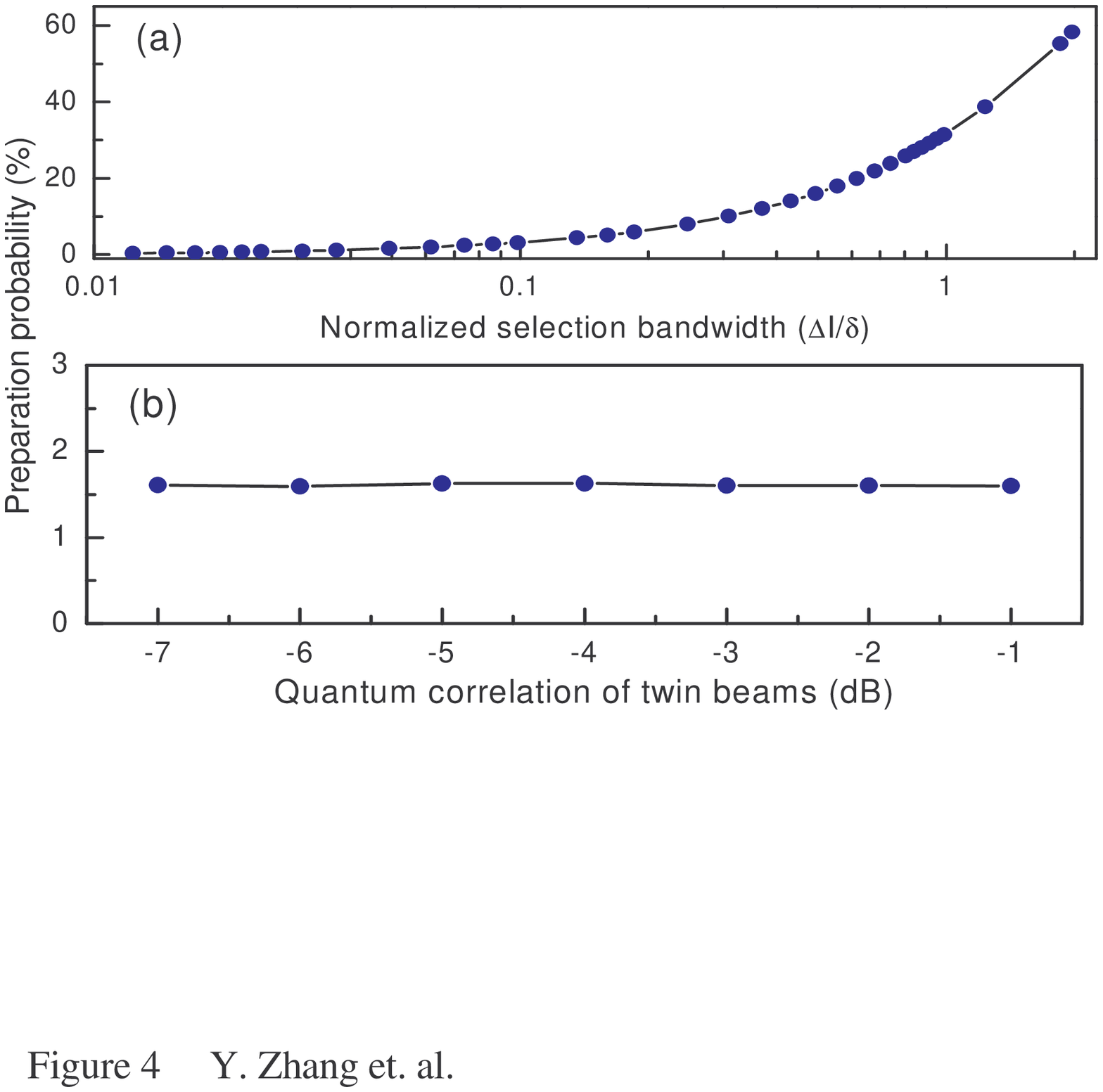}{%
\special{language "Scientific Word";type "GRAPHIC";maintain-aspect-ratio
TRUE;display "USEDEF";valid_file "F";width 3.5042in;height 2.418in;depth
0pt;original-width 7.7574in;original-height 11.0627in;cropleft
"0.0460";croptop "0.8566";cropright "0.9425";cropbottom "0.4245";filename
'fig4.ps';file-properties "XNPEU";}}

In conclusion, we achieved a conditional quantum correlation transfer using
two pairs of quantum-correlated twin beams initially produced from two OPOs.
We studied the influence of the selection bandwidth and noise reduction in
the twin beams on the transfer fidelity and preparation probability. This
experiment showed that the nonclassical features were transferred using the
method of conditional measurement even in a continuous-variable regime. When
two OPOs operating at different wavelength ranges are employed, our simple
protocol could be used to transfer the quantum correlation between two pairs
of twin beams having different wavelengths. The transfer is obviously not
restricted to the case of two pairs of twin beams. Using three or more pairs
of twin-beam sources, it would be possible to generate multipartite beams
with quantum correlation. The experimental protocol described here should
thus contribute to the future quantum information and communications in the
continuous variable regime.


\begin{thebibliography}{99}
\bibitem{kimble} A. S. Parkins and H. J. Kimble, J. of Opt. B \textbf{1},
496 (1999).

\bibitem{liu} C. Liu, et al., Nature \textbf{409}, 490 (2001).

\bibitem{kumar1} J. M. Huang and P. Kumar, Phys. Rev. Lett. \textbf{68},
2153 (1992).

\bibitem{furusawa} A. Furusawa, et al., Science \textbf{282}, 706 (1998); T.
C. Zhang et al., Phys. Rev. A \textbf{67}, 033802 (2003); W. P. Bowen et
al., Phys. Rev. A \textbf{67}, 032302 (2003); H. Yonezawa, et al., Nature 
\textbf{431}, 430 (2004).

\bibitem{heidmann} A. Heidmann, et al., Phys. Rev. Lett. \textbf{59}, 2555
(1987); J. Gao, et al., Opt. Lett. \textbf{23}, 870 (1998); K. Hayasaka, et
al., Opt. Lett. \textbf{29} 1665 (2004).

\bibitem{zhang1} Y. Zhang, et al., Opt. Lett. \textbf{27}, 1244 (2002); Y.
Zhang, et al., Opt. Express, \textbf{11}, 14 (2003).

\bibitem{fabre1} J. Laurat, et al., Phys. Rev. Lett. \textbf{91}, 213601
(2003).

\bibitem{feng} S. Feng and O. Pfister, Phys. Rev. Lett. \textbf{92}, 203601
(2004).

\bibitem{fabre2} J. Laurat, et al., Phys. Rev. A \textbf{70}, 042315 (2004).

\bibitem{lvovsky} A. I. Lvovsky, et al., Phys. Rev. Lett. \textbf{87},
050402 (2001).

\bibitem{grangier} J. Wenger, et al., Phys. Rev. Lett. \textbf{92}, 153601
(2004).

\bibitem{nha} H. Nha and H. J. Carmichael, Phys. Rev. Lett. \textbf{93},
020401 (2004).

\bibitem{opa} T. Opatrny, et al., Phys. Rev. A \textbf{61}, 032302 (2000);
P. T. Cochrane, et al., Phys. Rev. A \textbf{65}, 062306 (2002).

\bibitem{foster} G. T. Foster, et al., Phys. Rev. Lett. \textbf{85}, 3149
(2000).

\bibitem{lvovsky2} S. A. Babichev, et al., Phys. Rev. Lett. \textbf{92},
047903 (2004).

\bibitem{fabre3} J. Laurat, et al., Phys. Rev. A \textbf{69}, 033808 (2004).

\bibitem{walls} D. F. Walls and G. J. Milburn, Quantum Optics (Springer
Verlag, Berlin, 1994).

\bibitem{kumar} G. M. D'Ariano, et al., Phys. Rev. A \textbf{58}, 636
(1998); M. Vasilyev, et al., Phys. Rev. Lett. \textbf{84}, 2354 (2000).

\bibitem{zhang3} Y. Zhang, et al., J. Opt. Soc. B \textbf{21}, 1044 (2004).

\bibitem{lam} W. P. Bowen, et al., Phys. Rev. Lett. \textbf{90}, 043601
(2003).
\end{thebibliography}
\end{document}